\documentclass[onecolumn,twoside]{article}
\usepackage{graphics}
\usepackage{epstopdf}
\usepackage[utf8]{inputenc}
\usepackage[T2A]{fontenc}
\usepackage[english]{babel}
\usepackage{graphicx,epsfig}
\usepackage{floatflt}
\usepackage {amssymb}
\usepackage {amsmath}

\newcommand{\hb}{\\ \hspace*{2ex}}

\begin{document}
\title{Fixed points of mapping of N-point gravitational lenses}
\author{A.T.\,Kotvytskiy$^{1}$, V.Yu.\,Shablenko$^{2}$, E.S.\,Bronza$^{3}$\\[2mm] %English only
\begin{tabular}{l}
 $^1$ Department of Theoretical Physics, V.N. Karazin Kharkiv National University,\hb
 Kharkiv, Ukraine,  {\em kotvytskiy@gmail.com}\\
 $^2$ Department of Theoretical Physics, V.N. Karazin Kharkiv National University,\hb
 Kharkiv, Ukraine,  {\em shablenkov@gmail.com}\\
$^3$ Faculty of Computer Science, Kharkiv National University of Radio Electronics,\hb
 Kharkiv, Ukraine,  {\em eugene.bronza@gmail.com}\\
[2mm]
\end{tabular}
% no tabular, if one affiliation only
}
\date{}
\maketitle
ABSTRACT.
In this paper, we study fixed points of N-point gravitational lenses. We use complex form of lens mapping to study fixed points. Complex form has an advantage over coordinate one because we can describe N-point gravitational lens by system of two equation in coordinate form and we can describe it by one equation in complex form. We can easily transform the equation, which describe N-point gravitational lens, into polynomial equation that is convenient to use for our research. In our work, we present lens mapping as a linear combination of two mapping: complex analytical and identity mapping. Analytical mapping is specified by analytical function (deflection function). We studied necessary and sufficient conditions for the existence of deflection function and proved some theorems. Deflection function is analytical, rational, its zeroes are fixed points of lens mapping and their number is from 1 to N-1, poles of deflection function are coordinates of point masses, all poles are simple, the residues at the poles are equal to the value of point masses.\\
We used Gauss-Lucas theorem and proved that all fixed points of lens mapping are in the convex polygon. Vertices of the polygon consist of point masses. We proved theorem that can be used to find all fixed point of lens mapping.
On the basis of the above, we conclude that one-point gravitational lens has no fixed points, 2-point lens has only 1 fixed point, 3-point lens has 1 or 2 fixed points. Also we present expressions to calculate fixed points in 2-point and 3-point gravitational lenses. We present some examples of parametrization of point masses and distribution of fixed points for this parametrization.\\[1mm]
% Key words: Ierarchical structure according to the AAA list,
% the transition is marked by ":"
% different branches are separated by ";"
% same-level items are separated by ","
%\\[1mm]
{\bf Keywords}: gravitational lensing: lens mapping, fixed points, deflection function; complex analysis.
\\[2mm]

%===========================================================================================
%
{\bf 1. Introduction}\\[1mm]
Gravitational lensing is a phenomenon of deflection of light ray in a gravity field (Bliokh\&Minakov,1989; Zakharov,1997;Schneider,1999). With gravitational lensing, star systems and planets in star systems can be found. Recently, astronomers have observed a large number of gravitational lenses. In addition to one-point lenses, lenses with more than two components were also detected. In this paper, we show that such objects can have fixed points. In physical terms, fixed point of gravitational lens is a point in a picture plane that has such property: if we place source in fixed point, one of images is in this point.\\[2mm]
{\bf 2. General information and formulation of the problem}\\[1mm]
An N-point gravitational lens can be described by means of the following equation (Zakharov,1997;Schneider,1999):%
\begin{equation}
\vec{y}=\vec{x}-\sum_{n}m_{n}\frac{\vec{x}-\vec{l}_{n}}{\left\vert \vec
{x}-\vec{l}_{n}\right\vert ^{2}}, \label{lens_equat}%
\end{equation}
where $m_{n}$ are dimensionless masses whose position in the plane of the lens
is determined by the normalized radius-vectors $\vec{l}_{n}$. It is plain,
that $\sum_{n}m_{n}=1$.

We denote the set of radius-vectors $\vec{l}_{n}$ as $\Lambda=\left\{
l_{i}|i=1,2,...,N\right\}  $. Vector equation (\ref{lens_equat}) specifies
single-valued mapping%
\begin{equation}
L:\left(  R_{X}^{2}\backslash\Lambda\right)  \rightarrow R_{Y}^{2},
\label{mapping}%
\end{equation}
from vector space $R_{X}^{2}$ to vector space $R_{Y}^{2}$.

We introduce Cartesian coordinates, that transforms $R_{X}^{2}$ and $R_{Y}%
^{2}$ spaces into coordinate planes. Coordinate planes $R_{Y}^{2}$ and
$R_{X}^{2}$ are source plane and image plane respectively. Source plane
$R_{Y}^{2}$ and image plane $R_{X}^{2}$ are often united and called picture
plane in astrophysical literature.

Mapping (\ref{mapping}) can be described by system of equations:%
\begin{equation}
\left\{
\begin{array}
[c]{c}%
y_{1}=(x_{1}-\sum\limits_{n=1}^{N}m_{i}\frac{x_{1}-a_{n}}{\left(  x_{1}%
-a_{n}\right)  ^{2}+\left(  x_{2}-b_{n}\right)  ^{2}})\\
y_{2}=(x_{2}-\sum\limits_{n=1}^{N}m_{i}\frac{x_{2}-b_{n}}{\left(  x_{1}%
-a_{n}\right)  ^{2}+\left(  x_{2}-b_{n}\right)  ^{2}})
\end{array}
\right.  , \label{coord}%
\end{equation}
where $\left(  a_{n},b_{n}\right)  $ are coordinates of point $C_{n}$ of
radius-vector $\vec{l}_{n}$ in plane $R_{X}^{2}$.

Analytical research of (\ref{coord}) was in (Kotvytskiy\&Bronza\&Vovk, 2016; Bronza\&Kotvytskiy, 2017; Kotvytskiy\&Bronza\&Shablenko, 2017), and quasianalytical method of image construction was offered in (Kotvytskiy\&Bronza, 2016).

A point of single-valued mapping $L$ is fixed, if each of point coordinates is
invariant of $L$.

We need to substitute $y_{1}=x_{1}$ and $y_{2}=x_{2}$ and into system of
equations (\ref{coord}) and solve it to find fixed points.%
\begin{equation}
\left\{
\begin{array}
[c]{c}%
\sum\limits_{n=1}^{N}m_{i}\frac{x_{1}-a_{n}}{\left(  x_{1}-a_{n}\right)
^{2}+\left(  x_{2}-b_{n}\right)  ^{2}}=0\\
\sum\limits_{n=1}^{N}m_{i}\frac{x_{2}-b_{n}}{\left(  x_{1}-a_{n}\right)
^{2}+\left(  x_{2}-b_{n}\right)  ^{2}}=0
\end{array}
\right.
\end{equation}
Mapping $L$ is surjective. Inverse mapping%
\begin{equation}
L^{-1}:R_{Y}^{2}\rightarrow\left(  R_{X}^{2}\backslash\Lambda\right)  ,
\end{equation}
is multivalued. If $A_{0}$ - is a fixed point of single-valued mapping $L$,
then image of its image, when the mapping is reversed, is not coincide with
it, but includes it.%
\begin{equation}
A_{0}\in L^{-1}\left(  L\left(  A_{0}\right)  \right)  .
\end{equation}
In this paper we study the set $\Xi$ of fixed points of mapping $L$.

We set the mapping $L$ in complex form for effective application of
mathematical tool.
\\[2mm]
{\bf 3. Complexification of lens mapping $L$}\\[1mm]
Let define mapping (\ref{coord}) in complex form. We introduce complex
structure for $R_{X}^{2}$ and $R_{Y}^{2}$ , that transforms them into complex
planes $%
%TCIMACRO{\U{2102} }%
%BeginExpansion
\mathbb{C}
%EndExpansion
_{z}$ and $%
%TCIMACRO{\U{2102} }%
%BeginExpansion
\mathbb{C}
%EndExpansion
_{\zeta}$ respectively.

We introduce new complex variables $z$ and $\zeta$ . Let%
\begin{equation}
\operatorname{Re}z=x_{1},\operatorname{Im}z=x_{2},\operatorname{Re}\zeta
=y_{1},\operatorname{Im}\zeta=y_{2}.
\end{equation}
New variables related to old ones as%
\begin{equation}
\left\{
\begin{array}
[c]{c}%
x_{1}=\frac{z+\overline{z}}{2}\\
x_{2}=\frac{z-\overline{z}}{2}%
\end{array}
\right.  and\left\{
\begin{array}
[c]{c}%
y_{1}=\frac{\zeta+\overline{\zeta}}{2}\\
y_{2}=\frac{\zeta-\overline{\zeta}}{2}%
\end{array}
\right.  ,
\end{equation}
Now system (\ref{coord}) can be written as%
\begin{equation}
\zeta=z-\sum_{n=1}^{N}m_{n}\frac{1}{\overline{z}-\overline{A_{n}}%
},\label{compl_eq}%
\end{equation}
where $\sum_{n=1}^{N}m_{n}=1$ and $A_{n}=a_{n}+ib_{n}$; $n=1,2,...,N$.

We introduce function $\omega=\sum_{n=1}^{N}m_{n}\frac{1}{\overline
{z}-\overline{A_{n}}}$ and call it deflection function. Function is complex
conjugated to $\omega$ and defined:%
\begin{equation}
w=\sum_{n=1}^{N}m_{n}\frac{1}{z-A_{n}}%
\end{equation}
Functions $\omega$ and $w$ contain all the information about N-point
gravitational lens. Except that it is convenient to use function $w$, rather
than $\omega$, for application of methods of geometric function theory.

We have:%
\begin{equation}
\zeta=z-\overline{w}\left(  \overline{z}\right)  =z-\overline{w\left(
z\right)  }.
\end{equation}
Or:%
\begin{equation}
\overline{\zeta}=\overline{z}-w\left(  z\right)  . \label{compl(w)}%
\end{equation}
Thus, N-point lens can be described not only by the system of equation
(\ref{coord}) but also by the single equation (\ref{compl_eq}). Mapping
(\ref{mapping}) can be written as%
\begin{equation}
L:\left(
%TCIMACRO{\U{2102} }%
%BeginExpansion
\mathbb{C}
%EndExpansion
_{X}\backslash\Lambda\right)  \rightarrow%
%TCIMACRO{\U{2102} }%
%BeginExpansion
\mathbb{C}
%EndExpansion
_{Y},
\end{equation}
mapping of complex plane $%
%TCIMACRO{\U{2102} }%
%BeginExpansion
\mathbb{C}
%EndExpansion
_{X}$ into complex plane $%
%TCIMACRO{\U{2102} }%
%BeginExpansion
\mathbb{C}
%EndExpansion
_{Y}$.

We can obtain equation (\ref{compl(w)}) in another way. We can use equation (\ref{lens_equat}) (Witt, 1990).
\\[2mm]
{\bf 4. Some properties of $\zeta=\zeta\left(  z\right)  $ and $w=w\left(
z\right) .$}\\[1mm]

\textbf{Statement 4.1. }Function $\zeta=\zeta\left(  z\right)  $ is not an
analytic function.

\textbf{Proof.} Derivative of $\zeta=\zeta\left(  z\right)  $
\[
\frac{\partial\zeta}{\partial\overline{z}}=\frac{\partial}{\partial
\overline{z}}\left(  z-\overline{w}\left(  \overline{z}\right)  \right)
=1-\frac{\partial\overline{w}}{\partial\overline{z}}\neq0
\]
is not identity equal zero, therefore $\zeta$ is not analytic function.

\textbf{Statement 4.2. }Deflection function $w=w\left(  z\right)  $ is an
analytic function.

\textbf{Proof.} Derivative of $w=w\left(  z\right)  $
\[
\frac{\partial w}{\partial\overline{z}}=\frac{\partial}{\partial\overline{z}%
}\left(  \sum_{n=1}^{N}m_{n}\frac{1}{z-A_{n}}\right)  =\sum_{n=1}^{N}%
m_{n}\frac{\partial}{\partial\overline{z}}  \frac{1}{z-A_{n}}
\equiv0
\]
is identity equal zero, therefore $w$ in an analytic function.

\textbf{Statement 4.3.} Deflection function $w=w\left(  z\right)  $ is:
\begin{itemize}
\item rational function, i.e. $w=\frac{A\left(  z\right)  }{B\left(  z\right)
},$ where $A\left(  z\right)  $ and $B\left(  z\right)  $ are polynomials;
\item the denominator is a degree $\deg B\left(  z\right)  =N$, the numerator
is a degree $\deg A\left(  z\right)  =N-1$;
\item leading coefficients of $A\left(  z\right)  $ and $B\left(  z\right)  $
are equal 1.
\end{itemize}

\textbf{Proof.} We reduce the sum to common denominator%
\begin{equation}
w=\sum_{n=1}^{N}m_{n}\frac{1}{z-A_{n}}.
\end{equation}
Denominator of deflection function $B\left(  z\right)  =\prod\nolimits_{n=1}%
^{N}\left(  z-A_{n}\right)  $ is a degree $\deg B\left(  z\right)  =N$ leading
coefficient equals 1. Numerator $A\left(  z\right)  =\sum_{n=1}^{N}%
m_{n}z^{N-1}+...$ is a degree $\deg A\left(  z\right)  =N-1$, leading
coefficient of $A\left(  z\right)  $ equals $\sum_{n=1}^{N}m_{n}=1.$

\textbf{Theorem 4.4.} Deflection function $w$ can be written in form:%
\begin{equation}
a)w=\frac{Q^{\prime}\left(  z\right)  }{Q\left(  z\right)  },
\end{equation}
where $Q\left(  z\right)  =\prod\nolimits_{n=1}^{N}\left(  z-A_{n}\right)
^{m_{n}}$;%
\begin{equation}
b)w=\frac{1}{\deg P\left(  z\right)  }\frac{P^{\prime}\left(  z\right)
}{P\left(  z\right)  }, \label{deflpolin}%
\end{equation}
where $P\left(  z\right)  $ is polynomial;

\textbf{Proof.} a) $w=\sum\limits_{n=1}^{N}m_{n}\frac{1}{z-A_{n}}=\\=\sum\limits_{n=1}^{N}\left(  m_{n}\frac{d}{dz}\left(  \ln\left(  z-A_{n}\right)  \right)
\right)  =\\=\sum\limits_{n=1}^{N}\frac{d}{dz}\left(  \ln\left(  z-A_{n}\right)
^{m_{n}}\right)  =\\=\frac{d}{dz}\left(  \ln\left(  \prod\limits_{n=1}%
^{N}\left(  z-A_{n}\right)  ^{m_{n}}\right)  \right)  =\frac{d}{dz}\left(  \ln
Q\left(  z\right)  \right)  =\frac{Q^{\prime}\left(  z\right)  }{Q\left(
z\right)  }.$

We note, that function $Q\left(  z\right)  $ is not a polynomial.

\textbf{Proof.} b)$w=\frac{Q^{\prime}\left(  z\right)  }{Q\left(  z\right)
}=\frac{d}{dz}\left(  \ln Q\left(  z\right)  \right)  =\\=\frac{d}{dz}\left(
\ln\left(  \prod\nolimits_{n=1}^{N}\left(  z-A_{n}\right)  ^{m_{n}}\right)
\right)  .$

Assume without loss of generality, that numbers $m_{n}$ are rational.

Let $m_{n}=\frac{p_{n}}{q_{n}}$, where $p_{n}$ and $q_{n}$ are natural numbers
and coprime integers.

We substitute that into equation and transform it. Whence, we have:%
\begin{equation}
w=\frac{d}{dz}\left(  \frac{1}{h}\ln\left(  \prod\nolimits_{n=1}^{N}\left(
z-A_{n}\right)  ^{\frac{p_{n}}{q_{n}}h}\right)  \right)  , \label{defl}%
\end{equation}
where $h=\prod\nolimits_{n=1}^{N}q_{n}$. Let $s_{n}=\frac{p_{n}}{q_{n}}h$.
Numbers $s_{n}$ are natural numbers.

After transformation (\ref{defl}) we have:%
\begin{equation}
w=\frac{d}{dz}\left(  \frac{1}{h}\ln {P}\left(z\right)  \right)=\frac{d}{dz}\left(  \frac{1}%
{h}P\left(  z\right)  \right)  =\frac{1}{h}\frac{P^{\prime}
}{P},
\end{equation}
where $P\left(  z\right)  =\prod\limits_{n=1}^{N}\left(  z-A_{n}\right)
^{s_{n}}$ is polynomial. But then%
\begin{equation}
w=\frac{1}{h}\frac{1}{h}\frac{P^{\prime}\left(  z\right)  }{P\left(  z\right)
}.
\end{equation}
As well, leading coefficients of $A\left(  z\right)  $ and $B\left(  z\right)
$ are equal 1 and leading coefficient of $P^{\prime}\left(  z\right)  $ equal
$\deg P\left(  z\right)  $. We have: $h=\deg P\left(  z\right)  $, i.e. we
have (\ref{deflpolin}). QED.

\textbf{Remark 1 (to theorem 4.4).} Polynomials $P\left(  z\right)  $ and
$P^{\prime}\left(  z\right)  $ have the same roots as $B\left(  z\right)
=\prod\limits_{n=1}^{N}\left(  z-A_{n}\right)  $, but with different multiplicity.

\textbf{Remark 2 (to theorem 4.4).} Then since function $\omega$ is complex
conjugate to $w$, we obviously have:%
\begin{align}
\omega & =\overline{w}=\overline{\left(  \frac{1}{h}\frac{P^{\prime}\left(
z\right)  }{P\left(  z\right)  }\right)  }=\frac{1}{h}\frac{\overline{P\left(
z\right)  }}{\overline{P\left(  z\right)  }}=\\
& =\frac{1}{h}\frac{\left(  \overline{P\left(  z\right)  }\right)  ^{\prime}%
}{\overline{P\left(  z\right)  }}=\frac{1}{h}\frac{\overline{P}^{\prime
}\left(  \overline{z}\right)  }{\overline{P}\left(  \overline{z}\right)
}.\nonumber
\end{align}
Function $\omega$, exactly as $w$, can be expressed in form of ratio of two
polynomials. Numerator of the ratio is a derivative of denominator up to an
unessential constant multiplier.

\textbf{Remark 3 (to theorem 4.4).} Polynomials $P\left(  z\right)  $ and
$P^{\prime}\left(  z\right)  $ are not coprime. Fraction $\frac{P^{\prime
}\left(  z\right)  }{P\left(  z\right)  }$ can be reduced. Polynomials
$P\left(  z\right)  $ and $P^{\prime}\left(  z\right)  $ are coprime, if and
only if $m_{n}=\frac{1}{N},n=1,2,...,N$.

\textbf{Theorem 4.6.} Poles of function $w$ are $A_{n}$ points, which are
coordinates of masses. All poles are simple. For any poles are always true:
pole residue equal normalized mass at that point. The sum of residues at
finite points into complex plane equal one. At infinity equal minus one.
\textbf{Proof.} Obviously.
 \\[2mm]
{\bf 5. Fixed points of a lens mapping}\\[1mm]
\textbf{Theorem 5.1.(About quantity)} By $n_{0}$ denote a quantity of a fixed
points of mapping $L:\left(
%TCIMACRO{\U{2102} }%
%BeginExpansion
\mathbb{C}
%EndExpansion
_{X}\backslash\Lambda\right)  \rightarrow%
%TCIMACRO{\U{2102} }%
%BeginExpansion
\mathbb{C}
%EndExpansion
_{Y}$, then $n_{0}:1\leq n_{0}\leq N-1$.

\textbf{Proof.} Fixed points of function $\zeta=z-\overline{w}\left(
\overline{z}\right)  $ are roots of equation $z=z-\overline{w}\left(
\overline{z}\right)  $, i.e. $\overline{w}\left(  \overline{z}\right)  =0$. We
have $w\left(  z\right)  =0$, if we complex conjugate it.

Therefore, we have $\deg P^{\prime}\left(  z\right)  =N-1$ from the
representation (\ref{deflpolin}). Hence, the number of zeroes of function $w$
with regard to multiplicity is $N-1$. Polynomial $P^{\prime}\left(  z\right)
$ can have multiple zeroes. The number of different zeroes of polynomial
$P^{\prime}\left(  z\right)  $ is from $1$ to $N-1$.

We have the theorem about distribution of a fixed points of mapping $L$.

\textbf{Theorem (main) 5.2.} (About distribution) Fixed points of mapping $L$
are in the convex polygon that consists of point masses.

\textbf{Proof.} We use Gauss-Lucas theorem: if $P$ is a polynomial with
complex coefficients, all zeros of $P^{\prime}$ belong to the convex hull of
the set of zeros of $P$.

By theorem 5.1, fixed points of mapping $L$ are zeroes of the function $w$. By
theorem 4.5 we have representation (\ref{deflpolin}).

Since $P\left(  z\right)  =\prod\limits_{n=1}^{N}\left(  z-A_{n}\right)
^{s_{n}}$, roots of $P^{\prime}\left(  z\right)  $,are in the convex polygon
that consists of set $\left\{  A_{n}\right\}  $, because of Gauss-Lucas theorem (Prasolov,2014;Davydov,1964).

\textbf{Theorem 5.3.} (of finding fixed points and its number) Fixed points of
mapping $L$ for $N\geq2$ are roots of:%
\begin{equation}
H\left(  z\right)  =\frac{P\left(  z\right)  }{\gcd\left(  P\left(  z\right)
,P^{\prime}\left(  z\right)  \right)  },
\end{equation}
their number $n_{0}=\deg H\left(  z\right)  ,$ and estimation $n_{0}:1\leq
n_{0}\leq N-1$ is achieved.

\textbf{Proof.} The polynomial $P\left(  z\right)  $ is divided by the
polynomial $\gcd\left(  P\left(  z\right)  ,P^{\prime}\left(  z\right)
\right)  $. Therefore $B\left(  z\right)  $ is a polynomial. Polynomial
$\gcd\left(  P\left(  z\right)  ,P^{\prime}\left(  z\right)  \right)  $ has
only multiple roots. Multiplicity of roots of $\gcd\left(  P\left(  z\right)
,P^{\prime}\left(  z\right)  \right)  $ is one less then multiplicity of
$P\left(  z\right)  $. Hence all roots of polynomial $H\left(  z\right)  $ are
different and $n_{0}=\deg H\left(  z\right)  $.

2-point gravitational lens has one fixed point.

In general situation the number of fixed points is $n_{0}=N-1$.

For $N>2$, we have only one fixed point, if and only if all point masses are
equal and located at the vortexes of regular polygon.

\textbf{Remark 4 (to theorem 5.3).} Fixed points are missing from point
gravitational lens.
\\[2mm]
{\bf 6. Examples}\\[1mm]
For 1-point lens we have deflection function%
\[
w=\sum\limits_{n=1}^{N}\frac{m_{n}}{z-A_{n}}=\frac{m_{1}}{z-A_{1}},
\]
where $m_{1}=1$ and $A_{1}=0$.

For 2-point lens we have deflection function%
\[
w=\frac{m_{1}}{z-A_{1}}+\frac{m_{2}}{z-A_{2}}%
\]
where  $A_{1}$ and $A_{2}$ are coordinates of point masses and $m_{1}+m_{2}=1$.

With $m_{1}=s,m_{2}=1-s$ and $s\in\left[  0,1\right]  $ we have%
\begin{align*}
z_{st}  & =A_{1}+\left(  A_{2}-A_{1}\right)  s
\end{align*}

For 3-point lens we have deflection function%
\[
w=\frac{m_{1}}{z-A_{1}}+\frac{m_{2}}{z-A_{2}}+\frac{m_{3}}{z-A_{3}}%
\]
where $A_{1},A_{2},A_{3}$ are coordinates of point masses and $m_{1}%
+m_{2}+m_{3}=1$.

We have an equation for fixed points%
\begin{gather*}
z^{2}+A_{2}A_{3}m_{1}+A_{1}A_{3}m_{2}+A_{1}A_{2}m_{3}-\\
-\left(  A_{2}m_{1}+A_{3}m_{1}+A_{1}m_{2}+A_{3}m_{2}+A_{1}m_{3}+A_{2}%
m_{3}\right)  z=0
\end{gather*}
\begin{figure}[h]
\resizebox{1.0\hsize}{!}
{\includegraphics{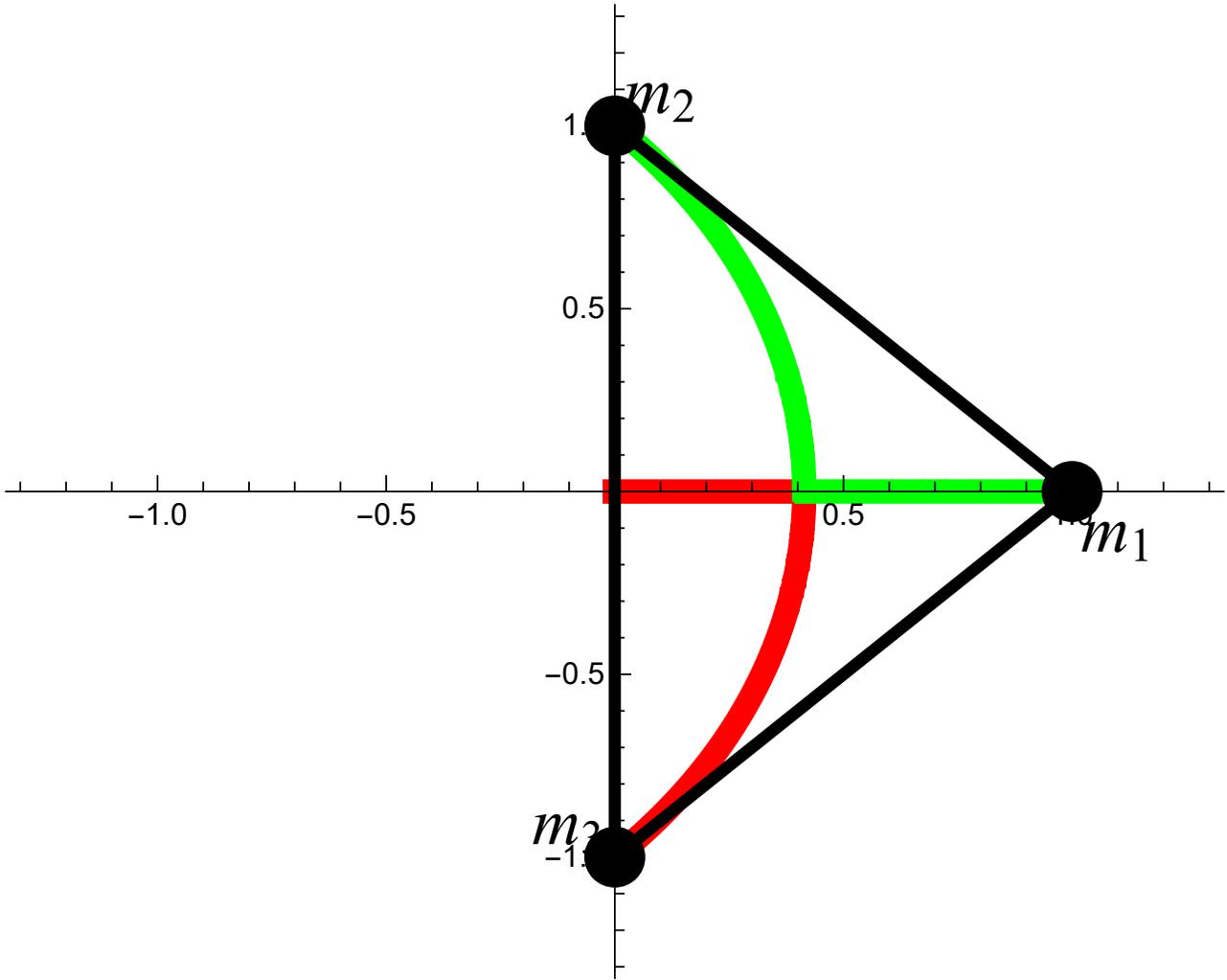}}
\label{hh}
\caption{3-point lens with $m_{1}=1-s,m_{2}=\frac{s}{2},m_{3}=\frac{s}{2},A_{1}=1,A_{2}=i,A_{3}=-i$}
\end{figure}
\begin{figure}[h]
\resizebox{1.0\hsize}{!}
{\includegraphics{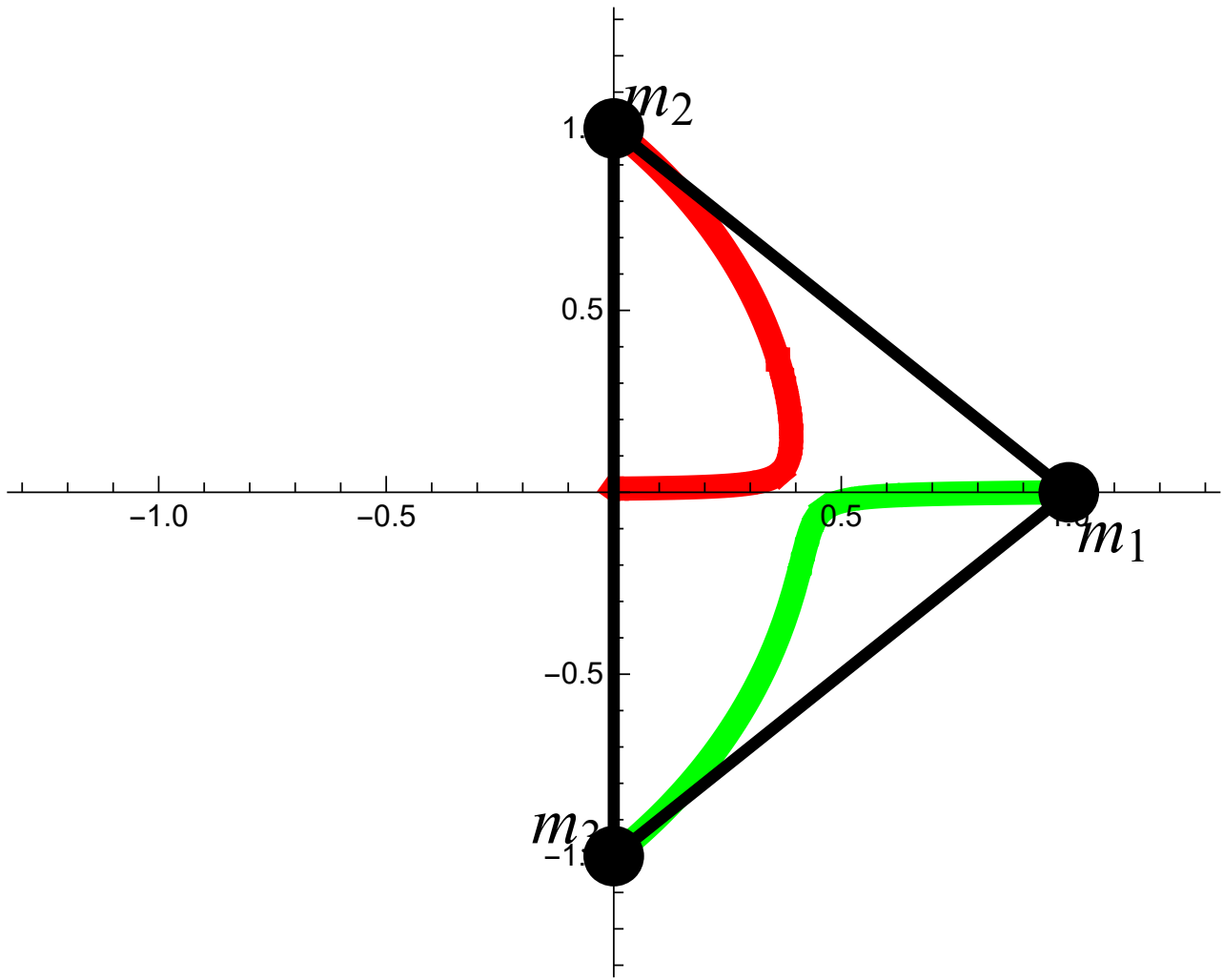}}
\label{hh}
\caption{3-point lens with $m_{1}=1-s,m_{2}=0.495 s,m_{3}=0.505 s,A_{1}=1,A_{2}=i,A_{3}=-i$}
\end{figure}
\begin{figure}[h]
\resizebox{1.0\hsize}{!}
{\includegraphics{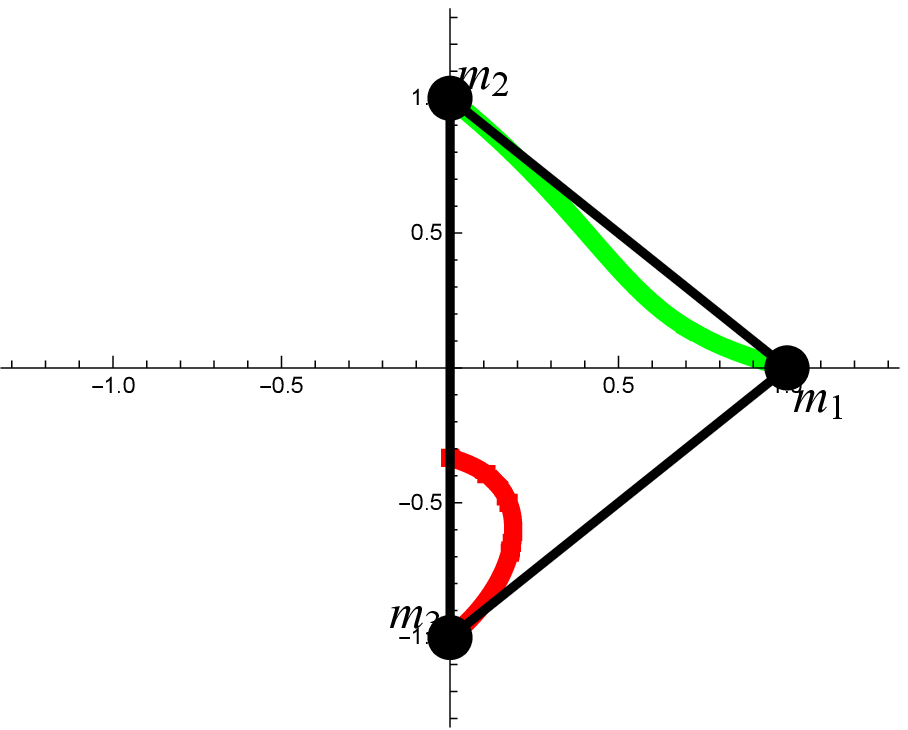}}
\label{hh}
\caption{3-point lens with $m_{1}=1-s,m_{2}=\frac{2s}{3},m_{3}=\frac{s}{3},A_{1}=1,A_{2}=i,A_{3}=-i$}
\end{figure}
\begin{figure}[h]
\resizebox{1.0\hsize}{!}
{\includegraphics{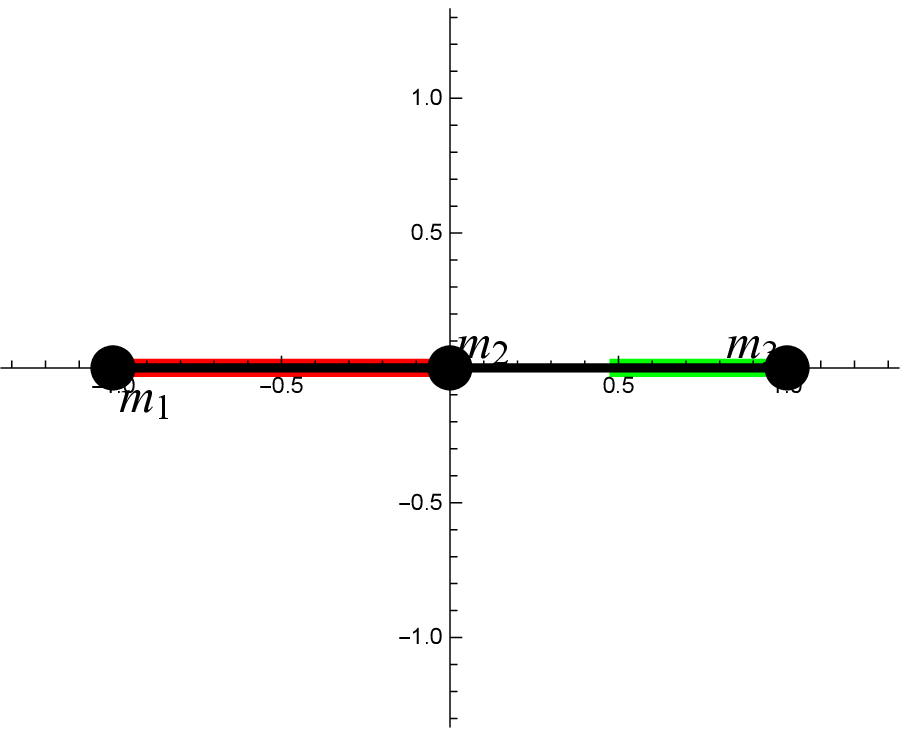}}
\label{hh}
\caption{3-point lens with $m_{1}=1-s,m_{2}=\frac{s}{2},m_{3}=\frac{s}{2},A_{1}=-1,A_{2}=0,A_{3}=1$}
\end{figure}
\\[2mm]
{\it Acknowledgements.} Albert Kotvytskiy and Volodymyr Shablenko thank for support SFFR, Ukraine, Project No. 32367.
\\[3mm]
{\bf References\\[2mm]}
Bliokh P.V., Minakov A.A.: 1989, Gravitational Lenses [in Russian]. (Naukova Dumka, Kiev), 240\\
Zakharov A.F.: 1997, Gravitacionnye linzy i mikrolinzy [in Russian]. (Janus-K, Moskow), 328\\
Schneider P., Ehlers J., Falco E.E.: 1999, Gravitational Lenses. (Springer-Verlag Berlin Heidelberg), 560\\
Kotvytskiy A.T., Bronza S.D., Vovk S.R.: 2016, Bulletin of Kharkiv Karazin National University “Physics”, {\bf 24}, 55 (arXiv:1809.05392)\\
Bronza S.D., Kotvytskiy A.T.: 2017, Bulletin of Kharkiv Karazin National University “Physics”, {\bf 26}, 6\\
Kotvytskiy A.T., Bronza S.D., Shablenko V.Yu.: 2017, Odessa Astronomical Publications, {\bf 30}, 35\\
Kotvytskiy A.T., Bronza S.D.: 2016, Odessa Astronomical Publications, {\bf 29}, 31\\
Witt H.J.: 1990, A\&A, {\bf 236},311\\
Praslov V.V.:2014, Polynomials[in Russian]. MCCME, 336\\
Davydov N.A.:1964, USSR Computational Mathematics and Mathematical Physics, {\bf 4},No.2, 257
\vfill
\end{document}